\documentclass{PoS}

\def\lsim{\raise0.3ex\hbox{$<$\kern-0.75em\raise-1.1ex\hbox{$\sim$}}}
\def\gsim{\raise0.3ex\hbox{$>$\kern-0.75em\raise-1.1ex\hbox{$\sim$}}}

\title{Universal O(N) scaling and the chiral critical line in
(2+1)-flavor QCD with small chemical potentials}

\ShortTitle{O(N) scaling and the chiral critical line}

\author{\speaker{Christian Schmidt\thanks{This work has
been supported in part by contracts DE-AC02-98CH10886 with the
U.S. Department of Energy, the BMBF under grant 06BI401, the
Gesellschaft f\"ur Schwerionenforschung under grant BILAER, the
Extreme Matter Institute under grant HA216/EMMI and the Deutsche
Forschungsgemeinschaft under grant GRK 881.  CS has been partially
supported through the Helmholtz International Center for FAIR which is
part of the Hessian LOEWE initiative.}
}\\Frankfurt Institute for Advanced
Studies, J.W.Goethe Universit\"at Frankfurt, D-60438 Frankfurt am
Main, Germany\\and\\GSI Helmholtzzentrum f\"ur Schwerionenforschung,
Planckstr.~1, D-64291 Darmstadt, Germany\\E-mail:
\email{cschmidt@fias.uni-frankfurt.de}}

\author{Swagato Mukherjee$^*$\\Physics Department, Brookhaven National
Laboratory,Upton, NY 11973, USA\\E-mail:
\email{swagato@quark.phy.bnl.gov}}

\abstract{We show that for small values of the chemical potential the
curvature of the phase transition line can be deduced from an analysis
of scaling properties of the chiral condensate and its
susceptibilities. We make use of a recent analysis of the magnetic
equation of state in (2+1)-flavor QCD where a connection between the
QCD parameters and the universal scaling fields could be
established. The remaining dependence of the reduced temperature on the
chemical potential can be fixed by an analysis of a mixed
susceptibility, obtained from a derivative with respect to quark mass
and chemical potential. We extract this dependence which describes the
curvature of the phase transition line, at two values of the cut-off,
$aT=1/4$ and $1/8$.  We find that cut-off effects are small for the
curvature parameter and determine the transition line in the chiral
limit to leading order in the light quark chemical potential.  We
obtain $T_c(\mu_B)/T_c(0) = 1 - 0.00656(66) (\mu_B/T)^2 +{\cal
O}(\mu_B^4)$.}

\FullConference{The XXVIII International Symposium on Lattice Field Theory, Lattice2010\\June 14-19, 2010\\Villasimius, Italy}

\begin{document}
\section{Introduction and Summary}
Extending lattice QCD calculations to non-zero baryon-chemical
potential or, equivalently, to non-zero net baryon number density is
known to be difficult in general. However, important information on
the QCD phase diagram can be deduced for small values of the chemical
potential by using well established numerical techniques such as
reweighting \cite{Fodor}, analytic continuation
\cite{analytic,Lombardo} or Taylor expansion \cite{Taylor,Gavai}.
At non-zero values of the chemical potential a phase boundary in the
temperature and chemical potential parameter space of QCD is well
defined only in the heavy quark limit or for vanishing quark
masses. In the former case the phase transition line corresponds to
the first order deconfinement transition in the pure gauge theory. At
infinite values of the quark mass this transition is independent of
the chemical potential and defines a straight line in the $T$-$\mu$
plane.  For a large range of quark mass values the transition line is
not unique.  It characterizes a region of (rapid) crossover in
thermodynamic quantities and a pseudo-critical temperature extracted
from these observables may differ somewhat, depending on the
observable that is used.  In the chiral limit, however, the transition
line is again well defined.  For sufficiently large strange quark mass
it defines a line of second order phase transitions in the
universality class of three dimensional $O(4)$ symmetric spin models
\cite{Pisarski}.

\begin{figure}
\begin{center}
\resizebox{0.49\textwidth}{!}{%
\includegraphics{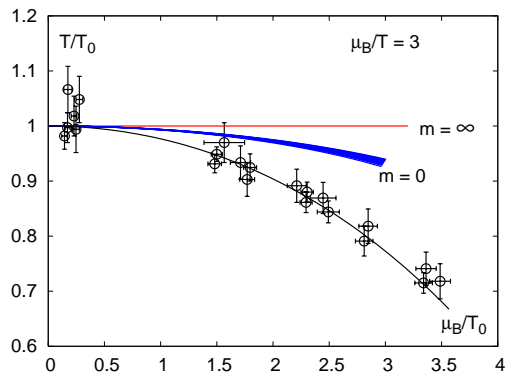}}
\end{center}
\caption{\label{fig:curv}
The curvature of the transition line in the ($T$,$\mu_B$)-diagram for
infinitely heavy quarks and -- as the result of this paper -- in the
chiral limit. For comparison we also show the freeze-out data from
heavy ion experiments, together with a parameterization for the
freeze-out line from \cite{cleymans}. Units are normalized to the 
transition
temperature at $\mu_B=0$ ($T_0$).}
\end{figure}
In a recent work \cite{our} we have shown that the curvature of the 
phase transition line in the chiral limit can be obtained from an analysis
of the universal scaling properties of a certain mixed susceptibility
which is defined by the leading order Taylor expansion coefficient of
the chiral condensate with respect to the light quark chemical potential.
Numerical calculations have been performed for (2+1)-flavor QCD
keeping the heavier strange quark mass close to its physical value and
decreasing the two degenerate light quark masses towards the massless
limit. We will make use of a recent scaling analysis
\cite{our, magnetic} of the chiral order parameter performed with an
improved staggered fermion action.  This study showed that the chiral
order parameter is well described by a universal scaling function
characteristic for a three dimensional, $O(N)$ universality class. 
As a result for the critical line in the chiral limit we find 
\begin{equation}
T(\mu_B)/T(0)=1-0.00656(66)(\mu_B/T)^2+{\cal O}((\mu_B/T)^4)\;.
\end{equation}
This curvature is about a factor two larger than the reweighting
results obtained in (2+1)-flavor QCD \cite{Fodor}. It is however
consistent with results obtained in calculations with imaginary
chemical potentials. In fact it lies in between the 2-flavor
\cite{analytic} and 3-flavor \cite{deForcrand3} simulations performed
with the standard staggered fermion formulation and also is consistent
with results reported from (2+1)-flavor simulations with imaginary
chemical potential performed with the action used also in this study
(p4-action) \cite{lat10}.  The result is most relevant for comparison
with the experimentally determined freeze-out curve as shown in
Fig.~\ref{fig:curv}. The parameterization found in \cite{cleymans} has
a curvature that is about a factor 3-4 larger which suggests that (for
sufficiently large values of the chemical potential) the freeze-out
line does not follow the critical line.
In the following two Sections we briefly review the scaling analysis of 
the order parameter and the mixed susceptibility.
 
\section{Magnetic equation of state}

In general we can separate two kinds of contributions to the free
energy density, a part ($f_s$) that will generate singularities in
higher order derivatives of the partition function and a regular
part ($f_r$), we have
\begin{equation}
f(T,m_l,m_s,\mu_q,\mu_s) = f_s(T,m_l,m_s,\mu_q,\mu_s) + 
f_r(T,m_l,m_s,\mu_q, \mu_s) \; .
\label{freeenergy}
\end{equation} 
In addition to the temperature $T$, light ($m_l)$ and strange ($m_s$)
quark masses we also allow for a dependence of the free energy density
on the quark chemical potentials.  Close to the chiral phase
transition temperature at vanishing chemical potential the singular
part $f_s$ will give rise to universal scaling properties of response
functions. This has been exploited to analyze basic universal features
of the QCD phase diagram close to criticality \cite{Hatta}.  Although
$f_s$ depends on many parameters of the QCD Lagrangian, the
universal behavior can be expressed in terms of only two relevant scaling
variables $t$ and $h$, that control deviations from criticality at
$(t,h)=(0,0)$.
%The singular part of the free energy density depends on the
%parameters of the QCD Lagrangian, e.g. the quark masses, and the
%external control parameters, temperature and chemical potentials,
%only through two relevant couplings. These scaling variables, $t$ and
%$h$, %control deviations from criticality, $(t,h)=(0,0)$, along the
%two relevant directions, which in the case of QCD characterize
%fluctuations of the energy and chiral condensate, respectively.
To leading order the scaling variable $h$ depends only on parameters
that break chiral symmetry in the light quark sector, while $t$
depends on all other couplings. In particular, $t$ will depend on the
light quark chemical potential while $h$ remains unaffected by this
in leading order,
\begin{eqnarray}
t &\equiv& \frac{1}{t_0}\left( \frac{T-T_c}{T_c} + 
\kappa_q \left( \frac{\mu_q}{T}\right)^2 
\right)
\ ,
\nonumber \\
h &\equiv& \frac{1}{h_0} \frac{m_l}{m_s} \ ,
\label{scalingfields}
\end{eqnarray}  
where $T_c$ is the phase transition temperature in the chiral limit
and $t_0$, $h_0$ are non-universal scale parameters (as is $T_c$).
While the combination $z_0=h_0^{1/\beta\delta}/t_0$ is unique for a
given theory, the values of $t_0$ and $h_0$ will change under
rescaling of the order parameter \cite{magnetic}. Just like
the transition temperature $T_c$ also $t_0$ and $h_0$ are
cut-off dependent and will need to be extrapolated to the continuum
limit.

The singular part of the free energy, $f_s$, is a homogeneous function
of its arguments. This can be used to rewrite it in terms of the
scaling variable $z=t/h^{1/\beta\delta}$ as
\begin{equation}
f_s(t,h) = h^{1+1/\delta} f_s(z,1) \equiv h^{1+1/\delta} f_s(z) \ .
\label{fs}
\end{equation}
where $\beta,\ \delta$ are critical exponents of the three dimensional
$O(N)$ universality class \cite{Engels2001}.

The universal critical behavior of the order parameter, 
$M\sim \partial f/\partial m_l$, 
is controlled by a scaling function $f_G$ that arises from
the singular part of the free energy density after taking a derivative
with respect to the light quark mass,
\begin{equation}
M(t,h) \;=\; h^{1/\delta} f_G(z) \; .
\label{order}
\end{equation}
The scaling function $f_G(z)$ is well-known for the $O(2)$ and $O(4)$
universality classes through studies of three dimensional spin models
\cite{engelsO2}.  This so-called magnetic equation of state,
Eq.~(\ref{order}), has been analyzed recently for (2+1)-flavor QCD using
an improved staggered fermion formulation (p4-action) on lattices with
temporal extent $N_\tau=4,8$ \cite{our, magnetic} and light quark
masses as small as $m_l/m_s=1/80$, which corresponds to a pion mass
that is about half its physical value. It could be shown that the
chiral order parameter can be mapped onto a universal $O(2)$ scaling
curve
\footnote{As we are working with staggered fermions $O(2)$ is expected to be the relevant universality class at finite lattice spacing. We find, however, that fits to the $O(4)$-symmetric model work equally well.} 
and the scale parameters $t_0,\ h_0,\ T_c$ could be extracted. 
The scaling analysis has been performed for two order parameters, that
are multiplicatively renormalized by multiplying the chiral condensate
with the strange quark mass, but differ in handling additive
divergences, linear in the quark mass\footnote{ At finite value s of
the cut-off these terms are, of course, finite and may be viewed as a
specific contribution to the regular part that will not alter the
scaling properties for sufficiently small values of the quark mass.},
\begin{eqnarray}
M_b &\equiv& \frac{m_s}{T^4} \langle \bar{\psi}\psi \rangle_l \; , \nonumber \\
M &\equiv&  \frac{m_s}{T^4} \left( \langle \bar\psi \psi \rangle_l -
\frac{m_l}{m_s} \langle \bar\psi \psi \rangle_s \right) \; .
\label{order_parameter}
\end{eqnarray}
All resulting fit parameters from fits with and without a regular
contribution are summarized in Table~\ref{tab:parameter}. Note that for 
$N_\tau=8$ only fits including a regular contribution have been possible, since the smallest available mass in this case has been $m_l/m_s=1/20$.

\begin{table}
\begin{center}
\vspace{0.3cm}
\begin{tabular}{|c|c|c|c|c|c|}
\hline
 $N_\tau$ & $M_i$ & $t_0$ & $h_0$ & $T_c(0)$~[MeV] & $z_0$ \\
\hline
\multicolumn{6}{|c|}{fit using the scaling term only}     \\
\hline
4 & $M_b$ & 0.0037(2)   & 0.0022(3)   & 194.5(4) & 6.8(5) \\
  & $M$   & 0.0048(5)   & 0.0048(2)   & 195.6(4) & 8.5(8) \\
\hline
\multicolumn{6}{|c|}{fit using scaling and regular terms}  \\
\hline
4 & $M_b$ & 0.00407(9)  & 0.00295(22) & 194.9(2) & 7.5(3) \\
  & $M$   & 0.00401(9)  & 0.00271(20) & 194.8(2) & 7.2(3) \\
8 & $M_b$ & 0.00271(21) & 0.00048(9)  & 174.1(8) & 3.8(5) \\
  & $M$   & 0.00302(22) & 0.00059(10) & 175.1(8) & 3.8(4) \\
\hline
\end{tabular}
\end{center}
\caption{Scale parameters determined from the scaling fits on lattices
of temporal extent $N_\tau=4$ and $8$. The last column 
gives $z_0\equiv h_0^{1/\beta\delta}/ t_0$. 
We give the results for parameters entering the definition of scaling 
functions for $M_b$ and the subtracted order parameter $M$ as defined 
in Eq.~(2.5). Only the former has been used in the analysis 
of the mixed susceptibilities. Note that fits including regular terms, give 
consistent determinations of the parameters of the scaling functions 
determined from $M_b$ and $M$, respectively.\label{tab:parameter}}
\end{table}

%When comparing results obtained for $N_\tau=4$ and $N_\tau=8$ one also
%has to take into account the dependence of the scale parameters on the
%strange quark mass. In fact, as the scaling analysis has been
%performed at bare strange quark mass values $m_s$, fixed in lattice
%units, the corresponding physical value in the chiral limit at $t=0$
%is only determined a posteriori, once $T_c$ has been determined. It
%turns out that the physical values of the strange quark mass in the
%$N_\tau=4$ and $8$ calculations differ at $T_c$ by about 10\%. One may
%account for this mismatch by reweighting the results for the chiral
%condensates in the light and strange quark masses
%\cite{unger}. However, we will not attempt to do this here.

%The main outcome of the $N_\tau=8$ scaling analysis, aside from
%confirming the good scaling properties of the order parameter at a
%twice smaller value of the lattice spacing, is a determination of the
%scale parameters %t_0,\ h_0$ and the transition temperature $T_c$ in
%the chiral limit, needed in the definition of the scaling variable
%$z$, {\it i.e.} the determination of $t_0,\ h_0$ and $T_c$. We
%summarize these results in Table~\ref{tab:parameter}.  In the next
%section we will make use of these scale parameters to determine the
%curvature of the phase transition line for small values of the quark
%chemical potential.

\section{Curvature of the critical line}
\label{sec:curve}
At leading order the light quark chemical potential only enters the
reduced temperature $t$, as introduced in Eq.~(\ref{scalingfields}).
Also at non-vanishing values of the quark chemical potential the phase
transition point is located at $t=0$.  The variation of the transition
temperature with chemical potential therefore is parameterized in terms
of the constant $\kappa_q$ introduced in Eq.~(\ref{scalingfields}),
\begin{equation}
\frac{T_c(\mu_q)}{T_c} = 1 -\kappa_q \left( \frac{\mu_q}{T} \right)^2
+{\cal O}\left(\left( \frac{\mu_q}{T}\right)^4\right) \; .
\label{criticalline}
\end{equation}
To determine the chiral phase transition line in the $T$-$\mu$ plane
we thus need to determine the proportionality constant $\kappa_q$.
This is, in fact, the only left over free parameter in universal
scaling functions that needs to be determined.  
%All other parameters
%($t_0,\ h_0,\ T_c\equiv T_c(\mu_q=0)$) have already been determined in
%the scaling analysis of the order parameter discussed in the previous
%section.

The constant $\kappa_q$ can be determined by analyzing the dependence
of the chiral condensate on the light quark chemical potential. To 
extract information about the
dependence of the scaling variable $t$ on $\kappa_q$ it suffices to
consider the leading order Taylor expansion coefficient of the chiral
condensate,
\begin{equation}
\frac{\langle \bar{\psi}\psi\rangle_l}{T^3} = \left( \frac{\langle
\bar{\psi}\psi\rangle_l}{T^3} \right)_{\mu_q=0} +
\frac{\chi_{m,q}}{2T} \left(\frac{\mu_q}{T}\right)^2 + {\cal
O}((\mu_q/T)^4) \; ,
\label{pbp_Taylor}
\end{equation}
where
\begin{equation}
 \frac{\chi_{m,q}}{T} = \frac{\partial^2 \langle
\bar{\psi}\psi\rangle_l/T^3}{\partial (\mu_q/T)^2} = \frac{\partial
\chi_q/T^2}{\partial m_l/T} \; .
\label{mixed}
\end{equation}
The mixed susceptibility $ \chi_{m,q}$ is proportional to the leading
order coefficient of the Taylor expansion of the chiral condensate,
which has been introduced in \cite{Ray,Taylor6}. It may also be viewed
as the quark mass derivative of the light quark number susceptibility
($\chi_q$). 
%Details of its definition in terms of inverses of the
%staggered fermion matrix and its derivatives with respect to the quark
%chemical potential are given in Appendix A of Ref.~\cite{Taylor6}.

%In the massless limit the chiral order parameter vanishes at $T_c$ and
%varies as $M\sim (-t)^{\beta}$.  Its derivative with respect to $t$
%thus will diverge at $T_c$ like $t^{\beta-1}$. The same singular
%behavior will thus show up in a derivative of the chiral condensate
%with respect to temperature as well as the second derivative with
%respect to $\mu_q/T$. The pre-factors of the singularity in ${\rm d}
%M/{\rm d}T$ and ${\rm d}^2 M/{\rm d}(\mu_q/T)^2$, however, will differ
%by a factor $2\kappa_q T_c$. We will make use of this relation to
%determine the curvature of the critical line at $\mu_q=0$.

In the vicinity of the critical point the mixed susceptibility can be 
expressed in terms of the scaling function 
$f'_G(z)\equiv {\rm d}f_G(z)/{\rm d} z$,  
\begin{equation}
\frac{\chi_{m,q}}{T} =  
\frac{2 \kappa_q T}{t_0 m_s} h^{-(1-\beta)/\beta\delta}  f'_G(z) \;  .
\label{mixedscaling}
\end{equation}
The scaling function $f'_G(z)$ is easily obtained from $f_G(z)$
by using the implicit parameterization for the latter given in 
Ref.~\cite{engelsO2}. We also note that $ \chi_{m,q}$ diverges as 
function of the light quark mass at $t=0$, {\it i.e.} at the chiral 
phase transition temperature. In contrast to
the chiral susceptibility, $\chi_m \sim \partial M/ \partial m_l$, which 
stays finite in the chiral limit only for $t>0$, the mixed susceptibility 
is finite for all $t\ne 0$.
For small values of the light quark mass numerical results for the 
mixed susceptibilities
$ \chi_{m,q}$ may be compared to the right hand side of 
Eq.~(\ref{mixedscaling}). Here all parameters that enter $f'_G(z)$ are
known and the only  undetermined parameter is $\kappa_q$. 
Using a subset of the data samples that have been used for the scaling
analysis of the order parameter \cite{our,magnetic}, we calculated the
mixed susceptibility $ \chi_{m,q}$ on lattices with temporal extent
$N_\tau=4$ for several values of the quark mass. For this analysis we
used data sets separated by 50 trajectories. For the lightest quark
mass ratio, $m_l/m_s=1/80$, we selected 4 and for the three heavier
quark mass ratios, $m_l/m_s=1/10,\ 1/20, \ 1/40$, we choose 6 values
of the gauge coupling in a narrow temperature interval close to the
chiral phase transition temperature $T_c$, {\it i.e.}  $-0.02\le
(T-T_c)/T_c \le 0.06$.  Typically this involved about 500 to 950 gauge
field configurations per parameter set, except for the lightest quark
mass ratio where we analyzed about 350 gauge field configuration.  On
each gauge field configuration we calculated the various operators
necessary to construct $ \chi_{m,q}$.

\begin{figure}
\begin{center}
\resizebox{0.49\textwidth}{!}{%
\includegraphics{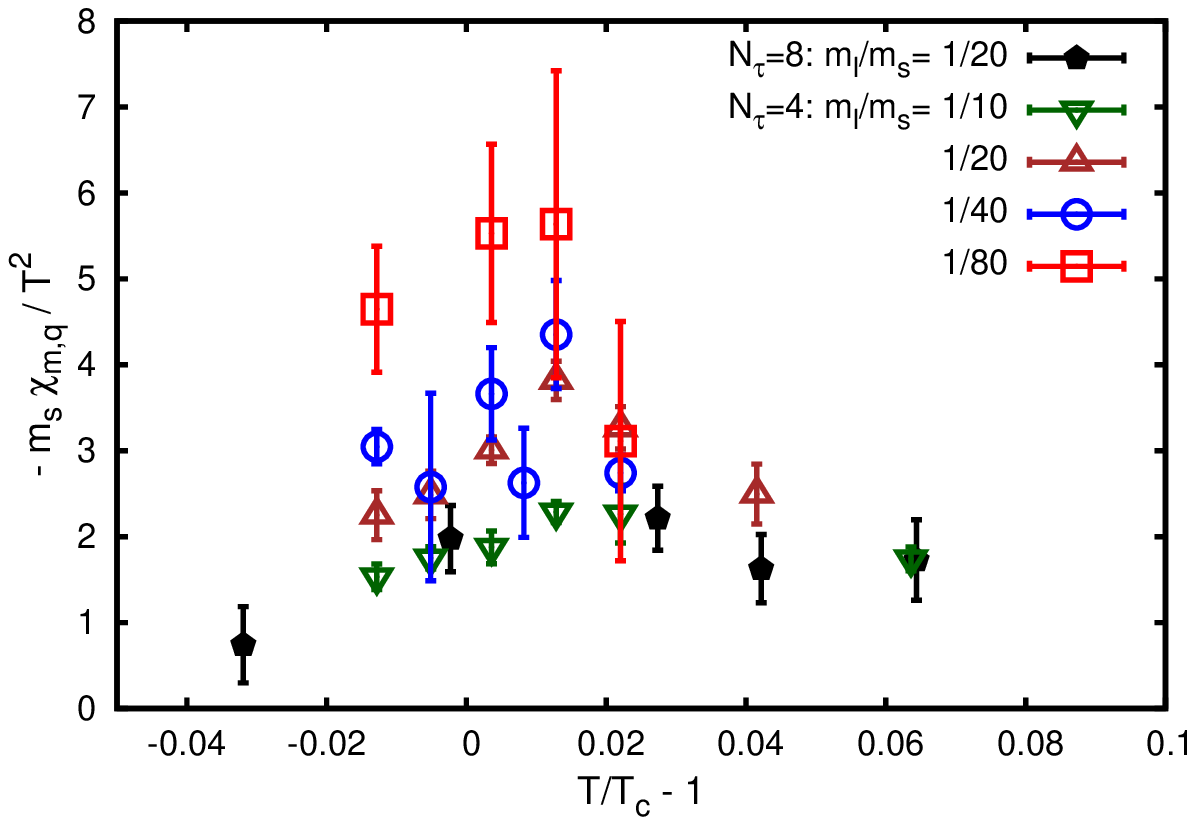}}
\resizebox{0.49\textwidth}{!}{%
\includegraphics{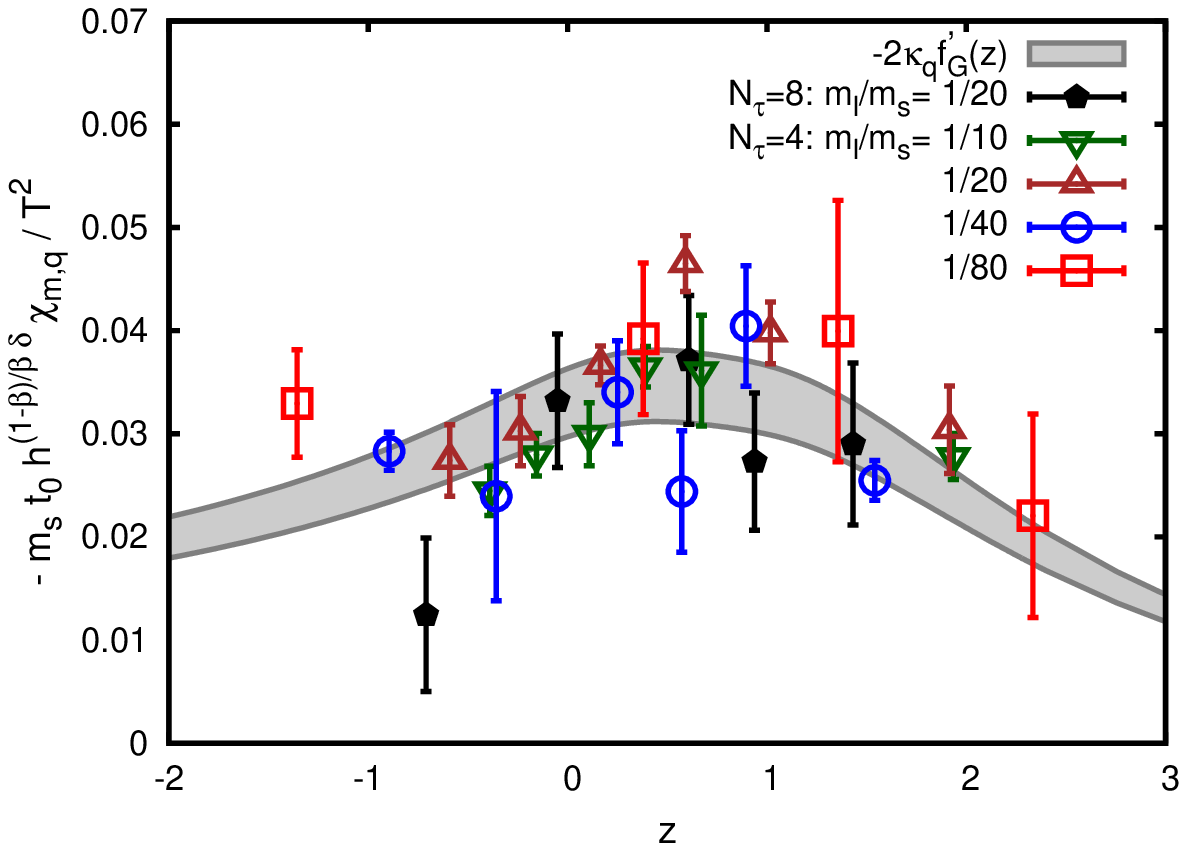}}
\end{center}
\caption{\label{fig:c2pbpiq}
The mixed light quark number susceptibility as a function of the
reduced temperature, $(T-T_c)/T_c$ (left) and the scaled mixed
susceptibility as function of the scaling variable
$z=t/h^{1/\beta\delta}$ (right). Shown are results obtained at two
values of the cut-off, $N_\tau =4$ (open symbols) and $N_\tau =8$
(filled symbols), and for several values of the light to strange quark
mass ratio. On the right hand side, the data is compared to the $O(2)$
scaling curve.}
\end{figure}

The calculation of the various operators required inversions of the
staggered fermion matrix with a large set of random noise vectors. We
used 500 noise vectors on each gauge field configuration and
constructed unbiased estimators for the various traces that need to be
calculated.  All these calculations could be performed very
efficiently on a GPU cluster.
 
Results obtained for the mixed light quark number susceptibility, 
$\chi_{m,q}$, on lattices with temporal extent $N_\tau=4$ are shown in
Fig.~\ref{fig:c2pbpiq} (left).  We clearly see that $ \chi_{m,q}$ increases
in the transition region with decreasing values of $m_l/m_s$.
Using the scaling relation given in Eq.~(\ref{mixedscaling}) we can
re-scale the data and obtain a unique scaling curve. This scaling
curve can be mapped onto the $O(2)$ scaling function $f'_G(z)$ with a
simple multiplicative rescaling factor, $2\kappa_q$.  The resulting
scaling plot is shown in Fig.~\ref{fig:c2pbpiq} (right).  To check for possible
contributions from scaling violating terms we have analyzed the data
separately for quark mass ratios $m_l/m_s = 1/10,\ 1/20$ and $m_l/m_s
= 1/40,\ 1/80$. These fits agree within statistical errors.  We then
determine the curvatures $\kappa_q$ from fits to the complete data
set. Results of these fits are summarized in Table~\ref{tab:fit}.
\begin{table}[t]
\begin{center}
\vspace{0.3cm}
\begin{tabular}{|c|c|c|c|}
\hline
 $N_\tau$ & $m_l/m_s$ & $\kappa_q$ & $\chi^2$/dof \\
\hline
4  & 1/10, 1/20 & 0.0598(26) & 3.5 \\
~  & 1/40, 1/80 & 0.0573(29) & 1.5 \\
\hline
8  & 1/20 & 0.0559(35) & 0.4 \\
\hline
4,\ 8  & all & 0.0591(17) & 2.1 \\
\hline
\end{tabular}
\end{center}
\caption{Determination of the curvature of the critical surface of the
chiral phase transition in $(2+1)$-flavor QCD as function of the light
quark chemical potential $\mu_q$. The table summarizes fits performed
separately for two lighter and two heavier quark mass sets as well as
the combined data set.\label{tab:fit}}
\end{table}
 
The scaling analysis performed for the mixed susceptibility on
lattices with temporal extent $N_\tau =4$ suggests that the
determination of the curvature parameter $\kappa_\mu$ can be reliably
performed with quark masses $m_l/m_s\lsim 1/10$. This is in accordance
with the scaling analysis of the order parameter itself
\cite{our,magnetic}. It thus seems to be safe to extract the curvature
parameter also at smaller values of the lattice spacing, {\it i.e.}
from our $N_\tau =8$ data set, by using the smallest quark mass ratio
available there, $m_l/m_s=1/20$.  We have performed calculations at
five values of the temperature using gauge field configurations on
$32^3\times 8$ lattices generated by the HotQCD collaboration \cite{eos}.  
For these parameter sets we have analyzed 300 to 600
gauge field configurations, which were separated by 100
trajectories. Again we used 500 noise vectors for the calculation of
all relevant operators on each of the gauge field configurations.  The
result of this analysis is shown in Fig.~\ref{fig:c2pbpiq} with filled
symbols. As can be seen they agree well with results obtained on
coarser lattices.

When rescaling data obtained for $ \chi_{m,q}$ to the $O(2)$ scaling
curve $f'_G(z)$ we need to take into account errors on the scaling
parameters $t_0$ and $z_0$ (or $h_0$). This leads to a 10\% error for
the determination of the curvature terms.  Performing a combined fit
to all results obtained for different quark mass values and lattice
spacings we obtain $\kappa_q = 0.059 (2)(4)$ or equivalently
$\kappa_B = 0.00656(22)(44)$.

\end{document}